\documentclass[journal]{IEEEtran}

\usepackage[colorlinks]{hyperref} 
\usepackage{fancyhdr}

\usepackage{graphicx}
 
\begin{document}

\title{Thermally Tunable Surface \\ Acoustic Wave Cavities}

\author{Andr\'e~Luiz~Oliveira~Bilobran, Alberto Garc\'ia-Crist\'obal, Paulo Ventura dos Santos, Andr\'es Cantarero
        and~Mauricio~Morais~de~Lima~Jr.
\thanks{A. L. O. Bilobran, A. Garc\'ia-Crist\'obal and M. M. de Lima Jr are with the Materials Science Institute, University of Valencia, P. O. Box 22085, E-46071 Valencia, Spain. A. Cantarero is with the Institute of Molecular Science, University of Valencia.  P. V. dos Santos is with the Paul-Drude-Institut f{\"u}r Festkörperelektronik, Hausvogteiplatz 5–7, D-10117 Berlin, Germany.  E-mail of correspondence: andre.oliveira@uv.es; mauricio.morais@uv.es.}

}


\pagestyle{fancy}
\fancyhf{}
\rhead{}

\onecolumn
© 2020 IEEE.  Personal use of this material is permitted.  Permission from IEEE must be obtained for all other uses, in any current or future media, including reprinting/republishing this material for advertising or promotional purposes, creating new collective works, for resale or redistribution to servers or lists, or reuse of any copyrighted component of this work in other works. DOI 10.1109/TUFFC.2019.2952982
\twocolumn
\maketitle

\begin{abstract}
We experimentally demonstrate the dynamical tuning of the acoustic field in a surface acoustic wave (SAW) cavity defined by a periodic arrangement of metal stripes on LiNbO$_3$ substrate. Applying a DC voltage to the ends of the metal grid results in a temperature rise due to resistive heating that changes the frequency response of the device up to 0.3\%, which can be used to control the acoustic transmission through the structure. The time scale of the switching is demonstrated to be of about 200 ms. In addition, we have also  performed finite element simulations of the transmission spectrum of a model system which exhibit a temperature dependence consistent with the experimental data. The advances shown here enable easy, continuous, dynamical control and could be applied for a variety of substrates.
\end{abstract}


\IEEEpeerreviewmaketitle

\section{Introduction}

\IEEEPARstart{S}{AW} propagating in periodic structures are the basis of a vast number of investigations.  SAW tags \cite{sawtagsreview2010}, for example, explore the possibility to use active or passive \cite{sawtagspassive2008} devices to encode information and use it in many applications, from traffic control, to security or identification of parts on conveyer lines, to name a few. Much similar designs are used for sensing applications, especially of temperature \cite{sawtempsensor2013} and mass \cite{sawmasssensor2013}. In addition, phononic crystals (PnCs) also explore the same architecture and have attracted a lot of attention lately. They are created by periodic arrays that vary the acoustic properties of the material, enabling the control of the propagation of elastic waves. The growing interest in this periodic structures is connected to their peculiar properties, such as anomalous dispersion (negative refraction) \cite{neg_refracted_index2012}, occurrence of band-gaps \cite{bandgap1993}, efficient wave-guiding \cite{waveguide2005} and near zero group velocity \cite{zerogroupvel1996}. Very recently, they were also used to enhance sensing capabilities \cite{tempsensor_pnc2019}.

Lately, a lot of effort were made to achieve the tunability of these devices. Among them some ideas are: use of variable impedances \cite{mauricioapl2012, crossapl1976, michiojjap2012, hashimotojjapl2013}, external magnetic field \cite{magnetic2003},  acoustoelectric interaction \cite{liapl2015, jpedrosapl2010}, electrostriction \cite{besanconapl2014, electrostcion2004} and thermal modulation \cite{dmitrievtp2007, wu2015}.

Here, we develop further on the last of these effects. We exploit the thermal instability of the substrate as a tool to control the acoustic field and demonstrate a continuous, memoryless and dynamical tunable device by which we can achieve 0.3\% frequency variation. We measure the time scale of the switching between operation modes to be of about $200$ ms. These features were not seen combined before in a simple device fabricated with standard lithography technique on an homogeneous substrate, as far as we know. Moreover, we devise ideas on how to intensify this effect, therefore expanding the usefulness of such framework.

\begin{figure}[ht]
\centering
\includegraphics[scale=0.5]{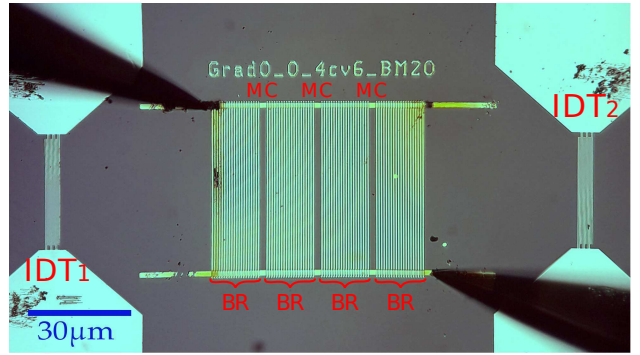}
\caption{Micrograph of sample with 3 microcavities(MC) defined by metal stripes forming Bragg reflectors (BR). The transmission (s$_{12}$ parameter) between IDTs 1 (source) and 2 (detection) is measured while a voltage is applied at the tips contacting the opposite ends of the metal grid.}
\label{micrograph}
\end{figure}
\subsection{SAW cavities device}
The SAW cavities are formed within the spacing between the distributed Bragg reflectors (BRs), each of which is composed of 20 gold stripes deposited on a  128$^{\circ}$-rotated Y-cut LiNbO$_3$ delay line. Each stripe reflects only a small part of the incoming wave. However, using $\lambda/4$ wide fingers and spacing them by  the same amount, results in the constructive interference of the reflected waves, thus creating efficient mirrors. Effectively, this creates a stop-band in the transmission spectrum of the incoming wave. The cavities are formed by properly spacing the BRs by an integer multiple of $\lambda/2$ , allowing the waves in between to interfere constructively, which in turn creates a transmission peak inside of the stop-band. We are going to focus specifically in one of these devices with three cavities (see Fig. \ref{micrograph}). The split-finger aluminum interdigital transducers (IDTs) of the delay lines were designed for an acoustic wavelength of $\lambda_{SAW} = 11.2\ \mu$m, which corresponds to a center resonance frequency of approximately 340 MHz for a Rayleigh wave propagating in the X-direction. Their metalization ratio is $1/2$ and they are formed by only 3 pairs of split-fingers, which makes them very broadband. This way, each of the resonances can be equally excited. The coupled cavities were fabricated by depositing NiCr/Au gratings with thicknesses $h=10/110$ nm within the SAW delay lines. 

\subsection{Related Work}
Recently, some of the authors demonstrated fundamental effects of quantum-wave transport \cite{mauricioprl2010} and acoustic field distribution control \cite{mauricioapl2012} by this type of structures. In both publications, as well as in this one, the cavities are built in such a way that they are coupled together. The limited reflectivity of the mirrors allow such behavior, acting as the acoustic equivalent of a Fabry-Perot resonator. The number of resonances is equal to the number of cavities (equal to 3 in Fig. \ref{micrograph}), each representing one Fabry-Perot mode. Due to the coupling the acoustic field can be activated in different regions of these multi-cavities device by exciting it with different eigenfrequencies. Investigations in which SAW are coupled to quantum emitters, for example, could take advantage of that \cite{quantdot2018}. Moreover, several works on sensors report an improvement on sensibility due to the use of coupled resonators \cite{coupledsensors2008}.

The tunability introduces a new degree of control by which one can, for example, switch from a high transmission state to a low transmission one (from a peak to a valley) while keeping the same frequency excitation.

\section{Results}
\subsection{Experiments}
\begin{figure}[ht]
\includegraphics[scale=0.38]{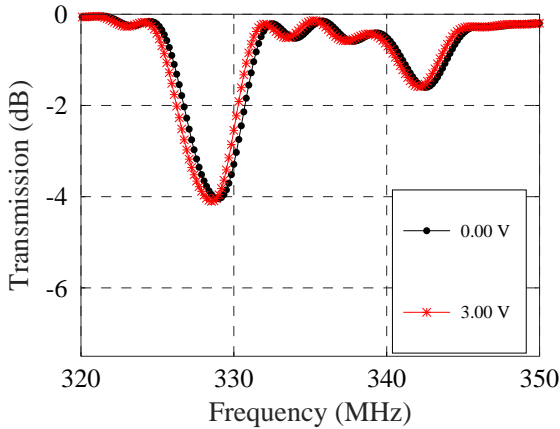} 
\caption{Measured s$_{12}$ (in dB) of the structure with 3 cavities. With a voltage applied to the grid ("x" markers) the frequency is shifted due to the heating of the substrate underneath caused by resistive heating of the metal stripes.}
\label{3cav_3V_compare}
\end{figure}

\begin{figure}[ht]
\includegraphics[scale=0.38]{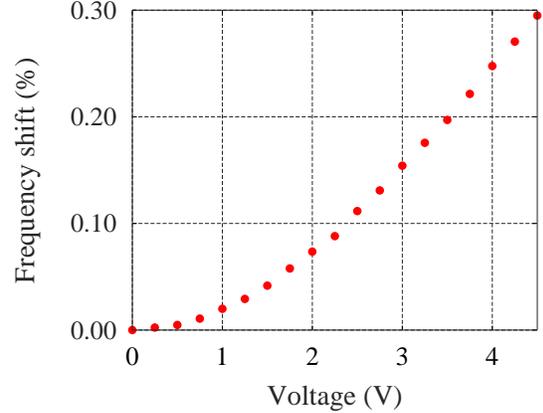} 
\caption{Frequency shift versus applied voltage of the first of the 3 peaks appearing in Fig. \ref{3cav_3V_compare}. }
\label{freq_shift}
\end{figure}

\begin{figure}[ht]
\includegraphics[scale=0.38]{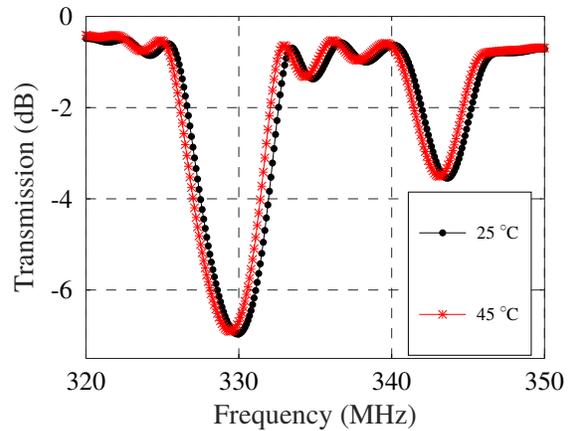} 
\caption{Simulated s$_{12}$ (in dB) of structure with 3 cavities. A FEM 2D model was built to simulate the acoustic field propagation along the sagittal plane of the structure.}
\label{3cav_sim_compare}
\end{figure}

By the use of shorted devices, i. e., where the ends of all stripes of the BR are connected, we contact the ends of the grid (see Fig. \ref{micrograph}) with a 5 $\mu$m copper tip and apply a DC voltage, $V_S$ on it. In Fig. \ref{3cav_3V_compare} one can see the frequency shift on the transmission spectrum of the device of three cavities for $V_S\ =\ 3\ V$ as compared with the response without applied voltage. The mechanism behind this effect is the resistive heating of the metal grid which warms up the sample. The temperature rise changes the relevant parameters of the substrate, such as the elastic and piezoelectric constants, resulting in the frequency shift that we observe. In this matter, the so-called temperature coefficient of frequency (TCF) is used to characterize the temperature stability of materials. The value for the 128$^{\circ}$-rotated Y-cut LiNbO$_3$ is of -76 ppm/$^{\circ}$C \cite{TCF1976}, which, for a 0.16\% frequency change corresponds to a temperature variation of 20 $^{\circ}$C approximately. 

The behavior of this shift according to the applied voltage can be seen in Fig. \ref{freq_shift}. It is worth comment that we could not apply higher voltages due to the fragility of the thin gold layers deposited. Nonetheless, we note that the shift could reach higher values if the grid were more extended in space, covering a large area of the delay line. Moreover, going to higher frequencies would also allow a higher contrast, since the wave would be more concentrated on the surface where the temperature rise is taking place.
\subsection{Simulations}
As a mean of verifying the consistency of this claim we have also performed finite element simulations where we include the temperature dependence of the material parameters. A 2D model was developed to mimic a cut containing the sagittal plane of the device. So we are considering the simulated structure infinite in the perpendicular direction, reducing thus the computation time. The reason why this simplification is suitable in this case is because we are focusing on Rayleigh waves, whose relevant displacements are within this plane.

The thickness of the model is around $1.6\lambda_{SAW}$ because SAWs' fields are known to be concentrated on a region whose depth is of the order of $\lambda_{SAW}$. In addition, a perfectly matching layer (PML) of depth $\lambda_{SAW}$ was added to absorb unwanted reflections. The minimum mesh size was $\lambda_{SAW}/25$, guaranteeing good convergence without demanding too much resources. In total there were approximately 270000 degrees of freedom (DOF) to solve for with a direct linear solver which coupled the solid mechanics and the electrostatics modules. A frequency domain study was used, taking 12 seconds to solve it for each specific frequency. The simulation was run on a desktop personal computer with Linux (Ubuntu) 3.40 GHz 64-bit processor and 16 GB RAM. All stripes on the BRs were simulated with the same height as in the experiments. However, the fingers of the IDTs are considered simply boundary conditions of zero thickness. The transmission between the two port device is calculated by applying floating electrical boundary condition to the fingers of the receiver IDT and taking the difference in voltage between each set of pairs. The parameters changes due to temperature were implemented according to the measurements of reference \cite{tempchange1971}.

As it can be seen in Fig. \ref{3cav_sim_compare}, the overall shape of the transmission, the difference between peaks and the shift due to the temperature rise, which are all the relevant physical parameters under consideration, are in excellent agreement with the experiment. The small discrepancies observed between the theoretical and experimental spectra can be partially attributed to limitations off the fabrication process. Roughness and inhomogeneities of the metal fingers impede the Bragg mirrors to reach its full reflection capacity, a feature not present on the simulation since the distance between strips can be made with arbitrary precision, as long as the mesh is fine enough.  The same argument holds for the transduction on the fingers of the IDT. In addition, the electrical boundary conditions are implemented in our model by forcing all stripes to be at zero potential. In reality, the non-zero resistivity of the metal results into departure of the propagation velocity of the ideal case. Moreover, in the samples, a small layer of NiCr is deposited under the gold stripes for better adhesion, a detail not included in our model. Another simplification of our model is related to the possibility of coupling of or conversion to different acoustic modes. As we are using a 2D model, a shear wave that might come into play is not taken into account. Nonetheless, we stress the fact that we use no adjustment parameters and the simple 2D model implemented is more than enough to predict all relevant features, even without the small corrections above mentioned.

\subsection{Time Scale}
We then proceed to inspect the time scale of this process. For that matter we apply a 50\% duty cycle square voltage, modulated from  0 V to 3 V.
\begin{figure}[ht]
\includegraphics[scale=0.38]{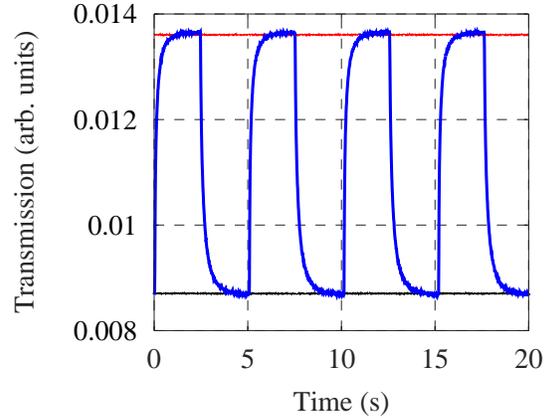} 
\caption{Measured s$_{12}$ parameter (linear scale, arbitrary units) as a function of time (thick solid line) for square modulated potential on the tips changing from 0 V to 3 V at a repetition rate of 0.2 Hz. The frequency is fixed at 332 MHz. The bottom and top solid horizontal lines are the transmission $s_{12}$ values for zero voltage and constant 3 V applied at the tips, respectively.}
\label{time_scale}
\end{figure}
By inspecting how the transmission changes for a specific frequency (zero span measurement), the acquisition of data was fast enough for us to probe the shift. Fig. \ref{time_scale} summarizes our findings. In it, three results are displayed: in black the transmission when no voltage is applied; in red, when a constant voltage of 3V is applied and in blue, when the square potential is on, all for a frequency of 332 MHz. By fitting an exponential decay $s_{12}=a\ $exp$(-t/\tau$), where $s_{12}$ is the transmission, $a$ is a constant and $\tau$ relates to the time scale of the process and averaging over several cycles we find $\tau_c \approx 220$ ms and $\tau_h \approx 160$ ms, where the subscripts $c$ and $h$ are related to the cooling and heating phases, respectively.

The units of the transmission here do not match the ones in Fig. \ref{3cav_3V_compare} due to the fact that, on the latter, a gating is performed on the raw data. We do a fast Fourier transform in order to filter the signals from cross-talk and reflections from other structures on the wafer. However, data on Fig. \ref{time_scale}  receive no such treatment because it is a zero span measurement.

\subsection{Discussion}
By comparing our simulations to the experiments we have demonstrated that a plausible raise in temperature is compatible to our findings. However, more effects can occur due to the application of an electrical potential which are, likewise, used as a tuning resource. One is electrostriction \cite{IEEEelectrostrictive2013}, which causes a piezoelectric stiffning or is also used to induce piezoelectricity in nonpiezoelectric substrates \cite{electrostrictive_transduction1975}. Another is simply the strain that  deforms the material, altering the path length of the wave. Nonetheless, approaches that explore one or both effects are fundamentally different from ours. They use the substrate as a capacitive mean. The voltage difference is applied either between top and bottom surfaces or on each of the IDTs, that is, on separated areas on the surface. Due to the dielectric properties of the substrate some have to use voltages on the order of kV. The strain caused by such an electric field distributed along the substrate is why its parameters are changed. In our case, however, the voltage is applied between different points of a piece of metal which sits on the surface. So the applied electric field is majorly confined to the metal. Moreover, some \cite{besanconapl2014, electrostcion2004} exhibit a memory effect. That is, the shift is seen to persist even after the perturbation is turned off. This makes them unsuitable for dynamical modulations like we show here, as it was already noted in \cite{JAPelectrostcion2007}. Others depend on the bias signal \cite{DCbias2005}, whereas ours is insensitive to that. These led us to conclude that the observed effect results majorly from the temperature rise.

In order to further support our claim that the observed effect is solely thermal we did an independent experiment in which no voltage was applied. On this set-up we mounted the wafer on top of a hot plate and measured the transmission through the structure with three cavities for different temperatures, waiting the appropriate time for the stabilization of the system. The data we observed is consistent with the previous results. For comparison, we note that the same 0.16\% shift that was observed when $3\ V$ was applied, in this new experiment corresponds to a 10 $^{\circ}$C rise above room temperature. This result, together with the arguments presented above, proves that we are dealing with a simple thermal effect. The temperatures were measured with an infrared thermometer aimed at the top of the surface of the wafer. According to its specifications, from a distance of around 3 cm, it reads the temperature of a spot of 2.5 mm diameter with a precision of $\pm$2 $^{\circ}$C. 

Contrasting our method to the ones available in the literature we see some improvements. As already mentioned, the ones relying on electro- \cite{besanconapl2014, electrostcion2004} or magnetostriction \cite{magnetic2003} exhibit an hysteresis curve. Acoustoelectric \cite{liapl2015, jpedrosapl2010} require more complex fabrication due to the multi-layer structure. And thermal modulation \cite{dmitrievtp2007, wu2015} do not mention dynamical tunability.

\section{Conclusion}
We have shown that the transmittance, $s_{12}$, of SAW coupled cavities can be dynamically tuned by temperature control of the substrate by means of an applied voltage. This scheme allows, for example, to switch from a high transmission state to a low transmission one, keeping the same frequency excitation. That is another degree of control of the acoustic field distribution of the devices, which could be used in various SAW-based technologies. The idea can be easily and universally implemented, taking advantage of the TCF of different materials.

\section*{Acknowledgment}

This project has received funding from the European Union’s Horizon 2020 research and innovation programme under the Marie Sklodowska-Curie Grant agreement No 642688 (SAWtrain).

\begin{IEEEbiographynophoto}{Andr\'e Luiz Oliveira Bilobran}
was born in Francisco Beltrao, Brazil. He received a Bachelor and a Master degree in Physics after attending the Federal University of Parana (Curitiba, Brazil), in 2013 and 2015, respectively. For his Master he developed a theoretical work on Foundation of Quantum Mechanics using tools of Quantum Information Theory. Before that, during the graduation, he worked in a project concerning friction of circular bodies.
\end{IEEEbiographynophoto}

\begin{IEEEbiographynophoto}{Alberto Garc\'ia-Crist\'obal.}
received the Bachelor and Ph.D. degrees in physics from the University of Valencia, Spain, in 1991 and 1996, respectively. He is currently a Professor with the University of  Valencia. His current research interests include the theoretical and numerical modeling of the interaction of electrons in semiconductor structures with optical and acoustic fields.
\end{IEEEbiographynophoto}

\begin{IEEEbiographynophoto}{Paulo Ventura dos Santos}
received his Master in Electrical engineering from the University of Campinas, Brazil and subsequently a Ph.D. degree in Physics from the University of Stuttgart, Stuttgart, Germany. After post-doctoral appointments at the Xerox Palo Alto Research Center in California, USA and at the Max-Planck Institute for Solid State Physics in Stuttgart, Germany, he joined the Paul-Drude-Institut für Festkörperelektronik in Berlin, Germany in 1997 as a senior scientist. His research interests include optical spectroscopy and mechanical properties of semiconductor nanostructures, including the interaction between semiconductor excitations with high-frequency acoustic waves. He is the author of approximately 300 publications, including peer-reviewed journal articles, book chapters, and patents.
\end{IEEEbiographynophoto}

\begin{IEEEbiographynophoto}{Andr\'es Cantarero}
is Full Professor in Condensed Matter Physics at the 
University of Valencia. After his Ph. D., he spent two years at the Max Planck 
Institute of Solid State Physics as a postdoc, where he came back several times as 
visiting professor. His main field of research has been optical spectroscopy studies 
of semiconductor nanostructures, mainly Raman spectroscopy. He developed several 
phenomenological models on Raman spectroscopy and lattice dynamics and, more 
recently, using density functional theory. He has also been working on transport 
and magnetic properties of solids. His main interests nowadays are thermoelectric 
materials, an exciting topic within energy harvesting. He organized several 
Schools and Conferences, is the co-author of around 330 publications, 3 book 
chapters and 3 patents. He was the principal investigator of more than 20 national 
projects, three European projects and several local projects. He belongs to the 
Molecular Science Institute of the University of Valencia and the Applied Physics 
Department.
\end{IEEEbiographynophoto}

\begin{IEEEbiographynophoto}{Mauricio Morais de Lima Jr.}
was born in Fortaleza, Brazil. He received his bachelor, M. Sc. and PhD degrees in physics from the University of Campinas (Unicamp), Campinas, Brazil, in 1996, 1998 and 2002, respectively. From 2002 to 2006, he has worked as a researcher on the Paul Drude Institute for Solid State Physics, Berlin, Germany. In 2006, he joined the Materials Science Institute of the University of Valencia as a ‘Ramón y Cajal’ fellow. Since then his major research line deals with the application of surface acoustic waves to the modulation of micro- and nano-structures.
\end{IEEEbiographynophoto}

\end{document}